\documentstyle[amssymb,multicol,epsf,aps]{revtex}
\renewcommand{\hbar}{h}

\newcommand{\r}{{\bf r}}
\newcommand{\bi}{\bibitem}

\newcommand{\Nh}{{\cal N}_h}
\newcommand{\Fc}{F_{\rm binaries}}

\newcommand{\Uc}{U_{\rm binaries}}
\newcommand{\Cc}{C_{\rm binaries}}
\newcommand{\Sc}{S_{\rm binaries}}
\newcommand{\Fh}{F_{\rm core}}
\newcommand{\Sh}{S_{\rm core}}
\newcommand{\Uh}{U_{\rm core}}
\newcommand{\Ch}{C_{\rm core}}
\newcommand{\Fr}{F_{\rm regular}}
\newcommand{\Cr}{C_{\rm regular}}
\newcommand{\Sr}{S_{\rm regular}}
\newcommand{\Ur}{U_{\rm regular}}

\newcommand{\St}{S_{\rm collaps}}
\newcommand{\Ut}{U_{\rm collaps}}
\newcommand{\Ct}{C_{\rm collaps}}

\newcommand{\BEQ}{\begin{equation}}
\newcommand{\EEQ}{\end{equation}}
\newcommand{\BEA}{\begin{eqnarray}}
\newcommand{\EEA}{\end{eqnarray}}

\renewcommand{\d}{{\rm d }}

\newcommand{\half}{\frac{1}{2}}

\newcommand{\nn}{\nonumber \\}
\begin{document}
\draft
\title
{Exact solution for core-collapsed isothermal star clusters}
\author{Th.~M.~Nieuwenhuizen}
\address{Department of Physics and Astronomy, University of Amsterdam
\\ Valckenierstraat 65, 1018 XE Amsterdam, The Netherlands}
\date{Version August 16, 1999; printout: \today}
\maketitle
\begin{abstract} 
Star clusters in isothermal spheres are studied from a thermodynamic
point of view. New density profiles are presented, 
that describe the collapsed phase at low temperatures.
At the transition a set of binaries is formed that carries 13 \% of 
the gravitational energy, while also a huge latent heat is generated.
The total energy of the binaries is fixed by thermodynamics. 
In the canonical ensemble all specific heats are positive, while
previously discussed negative specific heat solutions are metastable. 
In a microcanonical ensemble the latter remain partly dominant.
\end{abstract}
\hspace{1.7cm} keywords: ISM: globules, stellar dynamics, gravitation 
\begin{multicols}{2}
Globular star clusters typically have $10^5$-$10^6$
stars of solar mass. In our Galaxy about 150 of them
 have been observed, while their total 
number is estimated to be several hundreds.
Their most intriguing property is core collaps: Some 20 \% 
of the globulars in our Galaxy  have undergone 
 a ``violent relaxation'' towards a collapsed state~\cite{LyndenBell67}. 
The core collaps is stopped by the formation of 
strongly bound binary stars (``hard binaries''), 
and possibly also more exotic objects like
blue stragglers or black holes, that provide energy to the remaining
core~\cite{BinneyTremaine}
~\cite{MeylanHeggie}.

Self-gravitating star clusters confined 
to isothermal spheres have served as important models 
to understand the evolution of globular clusters. 
The problem was studied extensively by Chandrasekhar~\cite{Chandra}.
Todays status was presented in an overview by Padmanabhan~\cite{Paddy}. 
When varying the density contrast 
between center and boundary,  $d_0=n(0)/n(R)$,
there occur three regions:
{\it i}) the behavior is regular for $d_0\le 32.2$; 
{\it ii}) for $32.2\le d_0\le 702.9$ the solutions have a negative
heat capacity; {\it iii}) 
for $d_0> 702.9$ the  solutions are unstable
~\cite{Antonov}\cite{LyndenBellWood}. 

The occurrence of negative heat capacities is counter-intuitive, 
and has led to violent discussions.
Let us stress that we are discussing an equilibrium situation, 
which, in principle, should behave normally.
Out of equilibrium the situation would be different.
Negative heat capacities do occur for black holes
~\cite{Nblackhole} and even for simple model glasses~\cite{Nhammer},
without any contradiction of the {\it non-equilibrium thermodynamics}
~\cite{Nblackhole}~\cite{Nhammer}. In a canonical  equilibrium 
description the heat capacity is proportional to the
energy fluctuations, and necessarily positive.
Thirring pointed out that this objection does not apply in
a microcanonical approach~\cite{Thirring}. 
Lynden-Bell and Lynden-Bell considered the problem of negative heat
capacities in a more general setting~\cite{LyndenBell2}. 
In our view this situation  has nevertheless remained unsatisfactory.

As a standard thermodynamic description could be given for black
holes~\cite{Nblackhole}, we have started to believe that 
thermodynamics applies to gravitation and cosmology in general. 
Here we will set as our goal the description of the isothermal star 
cluster from a standard thermodynamic point of view. 
By this we mean: consider the lowest free energy, the specific heat, etc.
when  adiabatically lowering the temperature, thus employing
a canonical ensemble.  
The number $N\gg 1$ of identical stars, the total
mass $M=Nm$ and the sphere radius $R$ are kept fixed. 

 The temperature and energy scales are set by 
\BEQ T_G=\frac{GMm}{R}\qquad U_G=\frac{GM^2}{R}\EEQ
Let us introduce the reduced temperature
\BEQ \tau=\frac{2T}{T_G}=\frac{2RT}{GMm}\EEQ

The lowest temperature for which solutions are known is
 $T_c=0.396741\, T_G$~\cite{Paddy}, corresponding to
 $\tau_c=0.79346$. We shall address the
open question how the system behaves at lower $T$.
It should be described by the Poisson equation 
for the gravitational potential
 $\phi(\r)=-Gm\int\d^3r'\,n(\r')/|\r-\r'|$. For isothermal spheres
one has $n(\r)=Ae^{-\beta m\phi(\r)}$, leading to
\BEQ\label{Poisson}
 \nabla^2 \phi(\r)=4\pi Gm\, n(\r)=4\pi Gm A\,e^{-\beta m\phi({\r})}
\EEQ
where $n$ is the number density and $\beta=1/T$.
The  regular solutions are spherically symmetric and
can be mapped on the Emden equation
\BEQ y''+\frac{2}{x}y'=e^{-y}\EEQ
with $y(0)=y'(0)=0$. They are well understood, see~\cite{Paddy}.

Chandrasekhar~\cite{Chandra} and later Lynden-Bell and Wood
 ~\cite{LyndenBellWood}
mention the singular solution 
\BEQ\label{LBWs} n(r)=\frac{N}{4\pi Rr^2}; 
\qquad \beta m\phi(r)= 2\ln\frac{r}{R}-2.
\EEQ
Though these authors do not pay much attention to it,
it will become the corner stone of our analysis.
The solution (\ref{LBWs}) has $\tau=1$, and thus only exists at $T=T_G/2$. 
To describe the physics at low $T$, we have asked ourselves:
Can eq. (\ref{LBWs}) be deformed continuously for $\tau\neq 1$ ? 
We have therefore considered singular solutions with only cylindrical symmetry
\BEQ n(r,\theta)=\frac{N}{4\pi Rr^2}\,\frac{e^{-h(\cos\theta)}}
{\int_0^1\d c\,e^{-h(c)}}\label{haloR}\EEQ
subject to the north-south symmetry $\theta\to\pi-\theta$, 
implying $h(c)=h(-c)$. 
We call this the ``core'' solution, as there is no way to
identify a  halo in our static sphere with identical masses.
The gravitational potential reads
\BEQ \beta m \phi(r,\theta)=2\ln\frac{ r}{R}
+h(\cos\theta)-\frac{2}{\tau}-\int_0^1\d c\,h(c)\EEQ
where $\tau$ is to be fixed by the form of the solution.
Without loss of generality we may require $\int_0^1\d c\,h(c)=0$. 
From (\ref{Poisson}) it follows that $h$ has to obey
\BEQ\label{geqn} 2+(1-c^2)h''(c)-2c\,h'(c)=
\frac{2}{\tau} e^{-I_0}\, e^{-h(c)} \EEQ
with $I_0=\ln\int_0^1\d c\,e^{-h}$. As it should, 
the case $h=0$ leads us back to eq. (\ref{LBWs}) and $\tau=1$.

It is advantageous to introduce a new variable 
by setting $c=\tanh t$, and a new function $x(t)$, by defining
\BEQ \label{gdec}
h(c)=x(t)+\ln(1-c^2)\EEQ
Some algebra shows that eq. (\ref{geqn}) simplifies to
\BEQ \frac{\d^2 x}{\d t^2}=\frac{2}{\tau} e^{-I_0}\,e^{-x(t)}\EEQ
This is just the equation for one-dimensional motion in 
an exponential potential. The general solution is
\BEQ \label{hsol} 
x=2\ln\cosh[b(t-t_0)]-\ln(b^2\tau)-I_0\EEQ
Symmetry around $c=0$ imposes $t_0=0$. The parameter $b$ 
follows from the behavior for $t\to\infty$ or $c\to 1$.
The latter can be  determined
by integrating (\ref{geqn}) from 0 to $c$, using $h'(0)=0$. 
Comparing with (\ref{gdec}) and (\ref{hsol}) then yields $b=1/\tau$.
The solution of eq. (\ref{geqn}) thus reads 
\BEA \label{hsol=}
h(c)=&2&\ln\left[\left(\frac{1+c}{1-c}\right)^{1/(2\tau)}+
\left(\frac{1-c}{1+c}\right)^{1/(2\tau)}\right]\nn 
&+&\log(1-c^2)+\ln(\tau/4)-I_0
\EEA
The constant follows finally by imposing $\int_0^1\d c\,h=0$, yielding
\BEA
I_0&=&2\Nh(\frac{1}{\tau})-2+\ln\tau 
\EEA
The normalization of the $h$-function (\ref{hsol=}),
\BEA 
\Nh(\nu)&=&\int_0^1\d c\ln\left[\left(\frac{1+c}{1-c}\right)^{\nu/2}+
\left(\frac{1-c}{1+c}\right)^{\nu/2}\right]\nn
&=&\int_0^\infty\frac{\d t}{\cosh^2 t}\ln2\cosh (\nu\,t)
 \EEA
cannot be expressed in elementary functions. It has the behavior 
$\Nh(\nu)=\ln 2+(\pi^2/24)\nu^2$ for small $\nu$;  
$\Nh(1)=1$, $\Nh'(1)=\half$, while 
$\Nh(\nu)\to \nu\ln2+\pi^2/(24\nu)$ for large $\nu$.
From the second equality it is easily seen that 
$\nu^3\Nh''(\nu)=\Nh''(1/\nu)$, which implies the symmetry
\BEQ \Nh(\frac{1}{\tau})=\frac{1}{\tau}\Nh({\tau})\EEQ

It is important to know when the solution is stable.
Let us consider angular variations of the form 
$n(\r)\to n(r,\theta)[1+ f(y)]$, where $y= 
{\rm atanh}(\cos\theta)/\tau$, and satisfying $\int\,\d^3r\,nf=0$.
For small $f$ the free energy is invariant to first order, 
while to second order it reads 
\BEA 
\delta F&=&-\frac{Gm^2}{2}
\int \d^3 r\,\d^3 r' \frac{nfn'f'}{|\r-\r'|}
+\frac{T}{2}\int \d^3r\,{nf^2}\nn
&=&\int 
\frac{\d^3r\,\d^3r' n'f'}{4\pi |\r-\r'|}
\left(-2\pi Gm^2 nf-\frac{T}{2}\nabla^2 f\right)
\EEA
with $n'=n(r',\theta')$ and $f'=f(y')$. 
For our profile (\ref{haloR}) the term between brackets 
leads to the Schr\"odinger equation 
\BEQ -f''(y)-\frac{2\tau}{\cosh^2y}f(y)=E\,f(y)\EEQ
For $\tau>1$ there is a bound state,
$f=A\sinh y /\cosh^\lambda y$ with $\lambda=(-1+\sqrt{1+8\tau})/2$.
Since $E<0$ such fluctuations would lower the free energy, thus
making the solution unstable.
This has a simple physical interpretation:
if $\tau>1$ the density diverges on the north-south axis;
this unnatural situation is now seen to be unstable. 
If no other instabilities occur, the solution 
(\ref{haloR}), (\ref{hsol=}) is thus stable for $0<\tau<1$.

Let us now see what is the physical meaning of the new solution.
When coming from high $T$ the system is, of course, in the
regular solution. It can be described in terms of
the Milne variables $(u,v)$,
with $v=\beta T_G=2/\tau$, that satisfy~\cite{Chandra}\cite{Paddy}
\BEQ \frac{v}{u} \,\frac{\d u}{\d v}=-\frac{u+v-3}{u-1}\EEQ
with initial condition $u(0)=3$. 
In the $u$-$v$ plane the solution spirals counter clockwise 
around the point $(1,2)$, corresponding to infinite density contrast $d_0$.

The entropy is generally defined as
\BEQ S=\int 
\d^3r\,n(r)\,(\frac{5}{2}-\ln n(r)\Lambda^3)\EEQ
where $\Lambda(T)\sim T^{-1/2}$ is the thermal wavelength.
We shall subtract the constant 
$S_0=N\ln[{\sqrt{2}\pi R^3}/{N\Lambda^3(T_G)}]$ from 
$S$ in order to condense our notation.

The high temperature phase then has the properties
\BEA \label{Fr=}
\frac{\Fr}{U_G}&=&
\frac{1}{v}\left(2-u-v+\ln u +\frac{3}{2}\ln\frac{v}{2}\right)\\
\label{Ur=}
\frac{\Ur}{U_G}&=&\frac{2u-3}{2v}\\
\label{Sr=}
\frac{\Sr}{N}&=&2u+v-\frac{7}{2}-\ln u-\frac{3}{2}\ln\frac{v}{2}\\
\label{Cr=}
\frac{\Cr}{N}&=&\frac{u(u+v-3)}{u-1}+u-\frac{3}{2}
\EEA
We can now identify the {\it binodal temperature} $T_B=\half T_G$
and the {\it spinodal temperature} $T_S=T_c$.
In the interval $(T_S,T_B)$ (i.e. $\tau_c<\tau<1$)
there are two stable phases: the regular one and the new core state.
To lower the free energy, the system may collaps at $T_B$.
The free and internal energy, the entropy and the specific heat
are then 
\BEA \label{Fh=} 
\frac{\Fh}{U_G}&=&-\frac{1}{2}
-\frac{5\tau}{4}\ln\tau-\frac{1-\tau}{2}\Nh(\tau) \\
\label{Uh=} \frac{\Uh}{U_G}&=&-\half+\frac{5\tau}{4}
-\frac{1+\tau}{2}\Nh(\tau)\\ 
\frac{\Sh}{N} &=&\frac{5}{2}(1+\ln\tau)-2\Nh(\tau)\label{Sh=} \\
\label{Ch=} \frac{\Ch}{N} &=&\frac{5}{2}-\Nh(\tau)-(1+\tau)\Nh'(\tau)
\EEA
When we lower $T$ the system can also stay in the undercooled regular
state, and undergo a core collaps at some transition point 
$T_t=\half \tau_t T_G$ between $T_S$ and $T_B$.
If it did not yet do so, the system {\it must} collaps at $T_S$, 
simply because there is no other option.

At each of these transition points the free energy 
is discontinuous, a truly uncomfortable feature. 
The solution to this paradox is to assume that 
{\it also binary stars are formed} with free energy bridging this gap.
The collaps can thus be seen as a phase separation of the regular
phase into a core and binaries. 
It is well known that only a few hard binaries can be involved. 
Indeed, a finite central mass fraction
would lead to a behavior $\beta m\phi\sim-R/r$, which is incompatible 
with the Poisson equation (\ref{Poisson}). We therefore assume that 
the entropy of the binaries can be neglected, 
even though they must have a large energy.
Using $\Uc$=$\Fc$ we obtain
from the difference between (\ref{Fr=}) and (\ref{Fh=}) 
\BEA \label{Uct=}\frac{\Uc(T_t^-)}{U_G}&=&
\frac{1-\tau_t}{2}\Nh(\tau_t)-\half\nn&+&
(2-u_t+\ln \tau_tu_t)\frac{\tau_t}{2}\EEA
where $u_t$ is the $u$-value corresponding to $v=2/\tau_t$.
The above expression  ranges from $-0.0765$ at $T_B$ 
to $-0.10193$ at $T_S$, and is roughly equal to 13 \% of 
the pre-collaps gravitational energy.

As usual at first order phase transitions,
there is also a latent heat
\BEA\frac{\Delta U(T_t)}{U_G}
&=&1+(u_t-3+\Nh(\tau_t)-\frac{1}{2}\ln u_t\tau_t)\tau_t\EEA
At $T_B$ this equals 0.4062, being no less than 60.6 \% of the
precollaps gravitational energy, while 
at $T_S$ it equals 0.22043, still 27.8 \%. 
These striking effects are induced by the strong binding
nature of the core, expressed by its $1/r^2$ density
profile (\ref{haloR}). It leads to a low value of the core
energy (\ref{Uh=}). This is the manifestation of Lynden-Bell's violent 
relaxation~\cite{LyndenBell67}, and it is violent indeed !
 
\vspace{1cm}
\begin{figure}[htb]
\label{chiplot}
\epsfxsize=7cm
\centerline{\epsffile{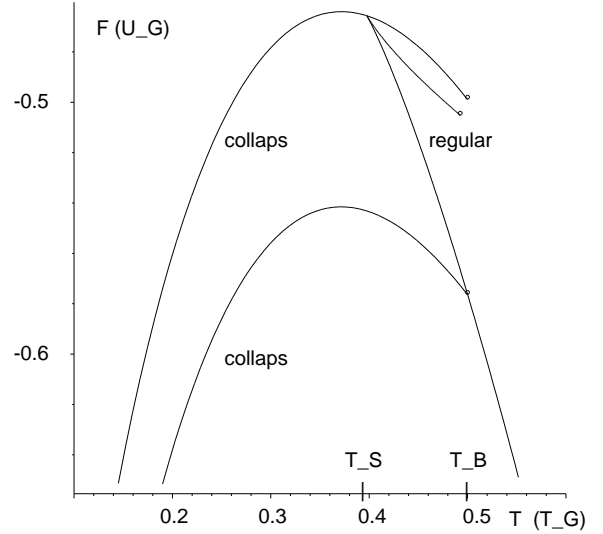}}
\vspace{-3cm}
\narrowtext
\caption{
Free energy $F$ as function of temperature $T$.
Below $T_B$ the system may enter the collapsed phase, 
which has lowest free energy. The regular phase may also be
undercooled; then collaps occurs at a lower temperature, and
 the modified binary energy shifts the collaps curve upwards. 
For the lowest possible collaps temperature, $T_S$,
the collaps free energy is presented by the upper curve.
The upper branch of the regular curve has 
a negative specific heat, and is metastable.
}\end{figure}

From eqs. (\ref{Sh=}) and (\ref{Ch=}) it is clear 
that  $\Ch\neq T\d\Sh/\d T$,
which looks quite disturbing.
However, we should not forget that the binary content 
is also a thermal part of the system. 
Assuming that equilibrium thermodynamics applies 
to the full system, we infer 
\BEQ \label{Fbinaries} 
\Cc-T\frac{\d \Sc}{\d T}=T\frac{\d \Sh}{\d T}
-\Ch\EEQ
Neglecting $\Sc$ and using  eqs.
(\ref{Sh=}), (\ref{Ch=}), this yields
\BEA \label{Cbinaries} \frac{\Cc}{N}
=\Nh(\tau)+(1-\tau)\Nh'(\tau)\EEA
It is positive,  implying that upon lowering $T$ 
more and more energy is stored in binaries.
Together with (\ref{Uct=}) this shows that 
{\it thermodynamics relates the energy of the binaries
to the properties of the core}.
Both contributions 
to the specific heat are positive,  and so is their sum
\BEA \label{Ccoll} \frac{\Ct}{N}
=\frac{5}{2} -2\tau \Nh'(\tau)\EEA
At $T_B$ one has $\Ch=N/2$, $\Cc=N$, while 
at $T_S$:  $\Ch=0.79504\,N$, $\Cc=0.99432\,N$.
Finally, at $T=0$ we find the Dulong-Petit type
behaviors $\Ch$ $=(5/2-\ln 2)N$, $\Cc=\ln 2\, N$.

The total internal energy of the collapsed phase reads
\BEA
\label{Ut=}\frac{\Ut}{U_G}&=&\frac{\tau_t}{2}(2-u_t+\ln\tau_tu_t)
-1+\frac{5\tau}{4}\nn
&&-\tau \Nh(\tau)-\int_{\tau}^{\tau_t}\d\nu\,\Nh(\nu)
\EEA
The final effect of metastability, leading to a $T_t<T_B$, is to
make the energy content of the binaries smaller.
In other words, knowledge of their energy allows the determination
of the collaps temperature $T_t$.

In Figure 1 we plot the total free energies of the problem.
Our approach solves the often discussed paradox 
of negative heat capacities of certain regular solutions.
Indeed, as shown in Fig. 1, for $T$ in a region above $T_B$
there are two regular solutions. The one with larger free energy
is also the one with the largest density contrast $(d_0>32.2)$, 
and a negative heat capacity. It is metastable and
cannot be reached by adiabatic cooling.

In Figure 2 we present the specific heat as function of
temperature for the regular solution, as well as for 
the binaries, the core and their sum. 

\begin{figure}[htb]
\label{Cplot}
\epsfxsize=7cm
\centerline{\epsffile{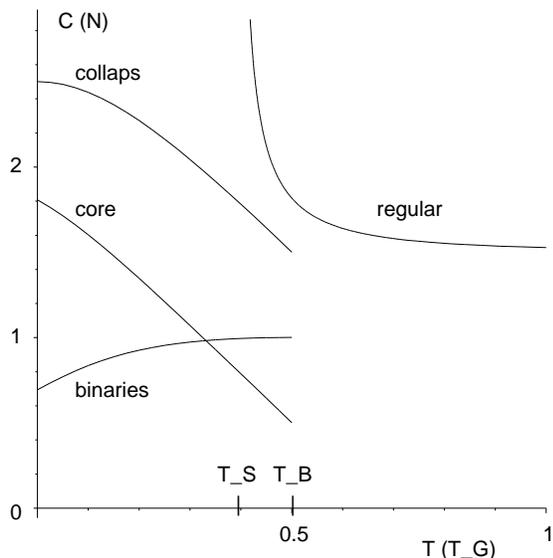}}
\vspace{-3cm}
\narrowtext
\caption{Specific heat $C$ as function of temperature $T$.
All contributions are positive. Upon cooling the system may collaps at
or below $T_B$. The largest chance occurs at $T_S$,
where  the regular solution has diverging energy fluctuations.}\end{figure}

Let us finally discuss the microcanonical situation.
We have to consider the entropy curves $S(U)$ on lowering $U$. 
For the regular solution it follows by eliminating $T$
between (\ref{Sr=}) and  (\ref{Ur=}).
The entropy of the stable collapsed phase has a maximum
$\St(\tau=1)=\half$. The regular phase has at that entropy already
a negative heat capacity, but that is not forbidden in the
microcanonical ensemble. For the energy of the collapsed phase we must
adjust onset value for the binaries, leading to the binodal value
$U_B=-0.2679\,U_G$.  The spinodal point is the point of instability
 of the regular solution, $U_S$ $=$ $-0.3342\,U_G$.
The microcanonical ensemble thus exploits the regular solutions more.
But, as seen from Figure 3, as soon as the
collapsed solutions are stable, they are also in the
 microcanonical approach the dominant ones.

In conclusion, we have presented new stable density profiles that describe
the core collapsed phase of isothermal clusters. 
Using standard thermodynamics the energy content stored in binaries
is determined. In the canonical ensemble all specific heats are positive.
Standard notions such as binodal and spinodal temperatures,
phase separation, undercooling and latent heat find their natural
place, just as in condensed matter physics.

Our results probably pertain to realistic clusters, whose
 core is to a good approximation isothermal.
Also there we expect that thermodynamics fixes the 
energy of the binaries. For clusters
with north-south axis perpendicular to our line of sight, the 
projected density of the profile (\ref{haloR}) is proportional to
$e^{-h(\cos\theta)}{\rm acos}(r/R)/r$.
Such a $1/r$ law is known from observations~\cite{Djorg}\cite{MeylanHeggie}.

\begin{figure}[htb]
\label{Splot}
\epsfxsize=7cm
\centerline{\epsffile{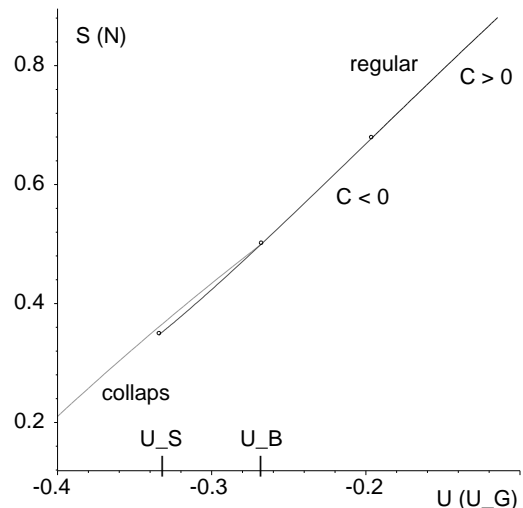}}
\narrowtext
\caption{Entropy $S$ as function of energy $U$.
Coming from high $U$, the system enters the 
region with negative heat capacity at $U_c=-0.1965\,U_G$.
At $U_B$ it can go to the collapsed state. 
If it fails to find it, it can ultimately collaps 
at $U_S$, where the regular solution becomes unstable.
}\end{figure}

 \acknowledgements
It is a pleasure to thank A. Allahverdyan for stimulating discussion.

\end{multicols}\end{document}